\documentclass[preprint,superscriptaddress]{revtex4-2}
\usepackage{graphicx}
\usepackage{bm}
\usepackage{amsmath,amsfonts,amssymb}
\usepackage[utf8]{inputenc}
\newcommand {\be} {\begin{equation}}
\newcommand {\ee} {\end{equation}}
\begin{document}

\title{Elastic proton-neutron and antiproton-neutron scattering in holographic QCD}

%---------------------------------------------------------------------------------------------------
\author{Akira Watanabe}
\email{watanabe@ctgu.edu.cn (Corresponding author)}
\affiliation{College of Science, China Three Gorges University, Yichang 443002, People's Republic of China}
\affiliation{Center for Astronomy and Space Sciences, China Three Gorges University, Yichang 443002, People's Republic of China}
%---------------------------------------------------------------------------------------------------
\author{Sayed Anwar Sirat}
\email{anwarsirat7@gmail.com}
\affiliation{College of Science, China Three Gorges University, Yichang 443002, People's Republic of China}
%---------------------------------------------------------------------------------------------------
\author{Zhibo Liu}
\email{202107020021014@ctgu.edu.cn}
\affiliation{College of Science, China Three Gorges University, Yichang 443002, People's Republic of China}
%---------------------------------------------------------------------------------------------------

\date{\today}

%%%%%%%%%%%%%%%%%%%%%%%%%%%%%%
\begin{abstract}
The total and differential cross sections of the elastic proton-neutron and antiproton-neutron scattering are studied in a holographic QCD model, focusing on the Regge regime.
Taking into account the Pomeron and Reggeon exchange, which are described by the Reggeized spin-2 glueball and vector meson propagator respectively, those cross sections are obtained.
It is presented that the currently available experimental data of the total cross sections can be well described within the model.
Once a single adjustable parameter is determined with the total cross section data, the differential cross sections can be calculated without any additional parameters.
Although the available differential cross section data are limited, it is found that our predictions are consistent with those.
\end{abstract}
%%%%%%%%%%%%%%%%%%%%%%%%%%%%%%

\maketitle

%%%%%%%%%%%%%%%%%%%%%%%%%%%%%%
\section{\label{sec:introduction}Introduction}
%%%%%%%%%%%%%%%%%%%%%%%%%%%%%%
High energy elastic hadron-hadron scattering has played an important role in improving our knowledge of the strong interaction for several decades.
Its cross sections shall, in principle, be described by QCD, but perturbative calculations are only available in some quite limited kinematic region, such as the high energy limit~\cite{Brodsky:1973kr,Matveev:1973ra,Lepage:1980fj}, due to the nonperturbative nature of the involved quark-gluon dynamics, which is the reason why in most kinematic regions effective approaches are basically needed.
The theoretical analysis for the forward scattering is especially difficult, since the underlying partonic dynamics is highly nonperturbative.
Historically, it is known that those cross sections can be well described by the Pomeron and Reggeon exchange, which can be interpreted as the multi-gluon and meson exchange, respectively.
Donnachie and Landshoff first showed that experimental data of various hadron-hadron total cross sections can be well reproduced with this description~\cite{Donnachie:1992ny}.
The Pomeron and Reggeon exchange can also be realized in the framework of holographic QCD, and in this work we apply this to the analysis of the elastic proton-neutron $(pn)$ and antiproton-neutron $(\bar{p}n)$ scattering.

Holographic QCD~\cite{Kruczenski:2003be,Son:2003et,Kruczenski:2003uq,Sakai:2004cn,Erlich:2005qh,Sakai:2005yt,DaRold:2005zs,Brodsky:2014yha} is an effective approach to QCD, which has been constructed based on the anti–de Sitter/conformal field theory (AdS/CFT) correspondence~\cite{Maldacena:1997re,Gubser:1998bc,Witten:1998qj}.
This method has served as a useful tool to perform analysis for the nonperturbative physical quantities, and has been successfully applied to studies of the spectrum and structure of hadrons~\cite{deTeramond:2005su,Karch:2006pv,Brodsky:2007hb,Abidin:2008ku,Abidin:2008hn,Abidin:2009hr,Branz:2010ub,Gutsche:2011vb,Li:2013oda,Gutsche:2017lyu,Lyubovitskij:2020gjz}.
Applications to various high energy scattering processes have also been intensively done, and a lot of successful results have been obtained so far~\cite{Polchinski:2001tt,Polchinski:2002jw,Brower:2006ea,Hatta:2007he,Pire:2008zf,Marquet:2010sf,Watanabe:2012uc,Watanabe:2013spa,Watanabe:2015mia,Watanabe:2018owy,Xie:2019soz,Burikham:2019zbo,Watanabe:2019zny,Liu:2022out,Liu:2022zsa}.
In particular, in Ref.~\cite{Liu:2022zsa} the cross sections of the elastic proton-proton $(pp)$ and proton-antiproton $(p \bar{p})$ scattering were investigated, considering the Pomeron and Reggeon exchange in holographic QCD, and it was shown that the experimental data can be well described in a wide kinematic region.

The present work is an extension of Ref.~\cite{Liu:2022zsa}, and we apply the framework to study the total and differential cross sections of the $pn$ and $\bar{p} n$ scattering, focusing on the Regge regime, in which the condition $s \gg |t|$ ($s$ and $t$ are the Mandelstam variables) is satisfied.
The Pomeron and Reggeon exchange are described by the Reggeized spin-2 glueball and vector meson propagator, respectively.
The Pomeron-nucleon couplings are described by the nucleon gravitational form factor, which can be obtained with the bottom-up AdS/QCD model~\cite{Abidin:2009hr}.
The model includes several parameters which should be determined with experimental data, but considering the universality of the Pomeron and Reggeon, part of the parameter set obtained in Ref.~\cite{Liu:2022zsa} can directly be employed in this study.

It is known that the isospin symmetry between the proton and neutron is a particularly good symmetry, which can be seen by the fact that the mass difference between these particles is surprisingly small.
However, since the charge difference affects the Reggeon-nucleon couplings, the Reggeon-neutron coupling constant needs to be newly determined, which can be made with the experimental data of the $pn$ total cross section.
Once this coupling constant is determined, the differential cross sections can be calculated without any additional parameters.
We present that the total cross section data can be well described within the model and our predictions for the differential cross section are also consistent with the data.
Since the Pomeron contribution becomes dominant as the energy increases, it is expected that the differences between the $pn$ and $\bar{p} n$ cross sections may decrease.
This behavior is explicitly shown for both the total and differential cross sections.

This paper is organized as follows.
In the next section, we introduce the model setup which describes the elastic $pn$ and $\bar{p} n$ scattering in the Regge regime.
We explain how to deal with the parameters involved in the model, and present our numerical results for the total and differential cross sections in Sec.~\ref{sec:numerical_results}.
Finally, the conclusion of this work is given in Sec.~\ref{sec:conclusion}.

%%%%%%%%%%%%%%%%%%%%%%%%%%%%%%
\section{\label{sec:model_setup}Model setup}
%%%%%%%%%%%%%%%%%%%%%%%%%%%%%%
In this study we investigate the elastic $pn$ and $\bar{p} n$ scattering in the Regge regime, taking into account the Pomeron and Reggeon exchange.
Following the preceding study~\cite{Liu:2022zsa}, here we introduce expressions for the total and differential cross section.
The Pomeron and Reggeon exchange can be described by the Reggeized spin-2 glueball and vector meson propagator, respectively.
Since we can consider these contributions separately, the total scattering amplitudes are written as
\be \label{eq:total_amplitude}
 \mathcal{A}_{\rm tot}^{pn(\bar{p}n)} = \mathcal{A}_{ g}^{pn(\bar{p}n)} + \mathcal{A}_{ v}^{pn(\bar{p}n)}.
 \ee

Considering the two-body scattering process, $1(p_1) + 2(p_2) \to 3(p_3) + 4(p_4)$, in the Regge limit the massive spin-2 glueball propagator is given by~\cite{Yamada:1982dx}
\be
D_{\alpha\beta\gamma\delta}^{g}(k) = \frac{-i}{2(k^2 + m_{g}^2)} (\eta_{\alpha\gamma}\eta_{\beta\delta} + \eta_{\alpha\delta}\eta_{\beta\gamma}),
\ee
where $k = p_3 - p_1$ and $m_{g}$ is the glueball mass.
The glueball-nucleon-nucleon vertex in the Regge limit is given by
\be
\Gamma_{g}^{\mu\nu} =
\frac{\lambda_{g} A(t)}{2} (\gamma^{\mu} P^{\nu} + \gamma^{\nu} P^{\mu}),
\ee
where $P = (p_1 + p_3)/2 $, $t=-k^{2}$, and $\lambda_{g}$ is the coupling constant which represents the strength of the glueball-nucleon coupling.
$A(t)$ is the nucleon form factor, which will be specified later.
The vector meson propagator and vector-nucleon-nucleon vertex are written as
\begin{align}
&D_{\mu\nu}^{v}(k) = \frac{i}{k^2 + m_{v}^2} \eta_{\mu\nu}, \\
&\Gamma_{v}^{\mu} = - i \lambda_{v} \gamma^{\mu},
\end{align}
respectively, where $m_{v}$ is the vector meson mass, and $\lambda_{v}$ is the coupling constant which represents the strength of the vector meson-nucleon coupling.

The scattering amplitudes for the glueball and vector meson exchange can be expressed as
\begin{align}
&\mathcal{A}_{g}^{pn(\bar{p}n)} = (\bar{u}_1 \Gamma_{g}^{\alpha\beta}u_3) D_{\alpha\beta\gamma\delta}^{g} (k) (\bar{u}_2 \Gamma_{g}^{\gamma\delta} u_4), \\
&\mathcal{A}_{v}^{pn(\bar{p}n)} = (\bar{u}_1 \Gamma_{v}^{\mu} u_3) D_{\mu\nu}^{v} (k) (\bar{u}_2 \Gamma_{v}^{\nu} u_4),
\end{align}
respectively, where $\bar{u}$ and $u$ are the nucleon spinors.
Since $s \gg |t|$ , $u_1 \approx u_3$, and $u_2\approx u_4$, the differential cross section can be obtained as
\begin{align} \label{eq:dcs}
\frac{d \sigma}{dt} &=
\frac{1}{16 \pi s^2} | \mathcal{A}^{pn(\bar{p}n)}_{\mathrm{tot}} |^2 \nonumber \\
&=\frac{\lambda_{g}^4 s^2 A^4 (t)}{16 \pi |t - m_{g}^2|^2} - \frac{\lambda_{g}^2 \lambda_{v}^2 A^2 (t) s}{8 \pi}\Bigg[\frac{1}{(t - m_g^2)^*}\times\frac{1}{t - m_v^2}+\frac{1}{(t - m_v^2)^*} \times \frac{1}{t - m_g^2}\Bigg] + \frac{\lambda_{v}^4}{4 \pi | t - m_{v}^2 |^2},
\end{align}
where the asterisk indicates complex conjugation.

Since this equation only describes the exchange of the lightest states, the both propagators need to be Reggeized to include contributions from the higher spin states on the Regge trajectories.
Following the Reggeization procedure explained in detail in Ref.~\cite{Anderson:2016zon}, the spin-2 glueball propagator needs to be replaced with
\be
\frac{1}{t - m_{g}^2} \rightarrow
\frac{\alpha'_{g}}{2} e^{- \frac{i \pi \alpha_{g} (t)}{2}} \left( \frac{\alpha'_{g} s}{2} \right)^{\alpha_{g} (t) - 2} \frac{\Gamma \left[3 - \frac{\chi_{g}}{2} \right] \Gamma \left[ 1 - \frac{\alpha_{g} (t)}{2} \right]}{\Gamma \left[2 - \frac{\chi_{g}}{2} + \frac{\alpha_{g} (t)}{2}\right]},
\ee
where $\chi_{g} \equiv \alpha_{g} (s) + \alpha_{g} (u) + \alpha_{g} (t)$ in which $\alpha_{g} (x) = \alpha_{g} (0) + \alpha'_{g} x$.
$\alpha_{g} (0)$ and $\alpha'_{g}$ correspond to the Pomeron intercept and slope, respectively.
On the other hand, the vector meson propagator needs to be replaced with
\be
\frac{1}{t - m_{v}^2} \rightarrow
\alpha'_{v} e^{-\frac{i \pi \alpha_{v} (t)}{2}} \sin \left[ \frac{\pi \alpha_{v} (t)}{2} \right] \left( \alpha'_{v} s\right)^{\alpha_{v} (t) - 1}\Gamma[- \alpha_{v} (t)],
\ee
where $\alpha_{v} (t) = \alpha_{v}(0) + \alpha'_{v} t$.

Therefore, the differential cross section which includes contributions from the higher spin states can be obtained as
\begin{align} \label{eq:dcs_final}
\frac{d \sigma}{dt} = 
&\frac{\lambda_{g}^4 s^2 A^4 (t)}{16 \pi} \left[ \frac{\alpha'_{g}}{2} \frac{\Gamma\left[3 - \frac{\chi_{g}}{2} \right] \Gamma \left[1 - \frac{\alpha_{g} (t)}{2} \right]}{\Gamma \left[2 - \frac{\chi_{g}}{2} + \frac{\alpha_{g} (t)}{2} \right]} \left( \frac{\alpha'_{g} s}{2}\right)^{\alpha_{g} (t) - 2} \right]^2 \nonumber \\
&- \frac{\lambda_{g}^2\lambda_{v}^2 s A^2 (t)}{4 \pi}\left[ \frac{\alpha'_{g}}{2} \frac{\Gamma \left[3 - \frac{\chi_{g}}{2} \right]\Gamma \left[1 - \frac{\alpha_{g} (t)}{2} \right]}{\Gamma \left[2 - \frac{\chi_{g}}{2} + \frac{\alpha_{g} (t)}{2}\right]} \left( \frac{\alpha'_{g} s}{2} \right)^{\alpha_{g} (t) - 2}\right] \nonumber \\
&\qquad \times \left[ \alpha'_{v} \sin \bigg( \frac{\pi \alpha_{v} (t)}{2} \bigg) (\alpha'_{v} s)^{\alpha_{v} (t) - 1}\Gamma[- \alpha_{v} (t)]\right] \cos \left[ \frac{\pi}{2} (\alpha_g (t) - \alpha_v (t)) \right] \nonumber \\
&+ \frac{\lambda_{v}^4}{4 \pi}\left[ \alpha'_{v} \sin \bigg( \frac{\pi \alpha_{v} (t)}{2} \bigg) (\alpha'_{v} s)^{\alpha_{v} (t) - 1} \Gamma[- \alpha_{v} (t)] \right]^2,
\end{align}
which leads to the total cross section:
\begin{align} \label{eq:tcs_final}
\sigma_{\mathrm{tot}} =
&\lambda_{g}^2 \sin \left( \frac{\pi \alpha_{g} (0)}{2} \right) \frac{\Gamma \left[3 - \frac{\chi_{g}}{2} \right] \Gamma \left[1 - \frac{\alpha_{g} (t)}{2} \right]}{\Gamma \left[2 - \frac{\chi_{g}}{2} + \frac{\alpha_{g} (t)}{2}\right]} \left( \frac{\alpha'_{g} s}{2} \right)^{\alpha_{g} (0) - 1} \nonumber\\
&- 2 \lambda_{v}^2 \alpha'_{v} \sin^2 \left( \frac{\pi \alpha_{v} (0)}{2} \right) (\alpha'_{v} s)^{\alpha_{v} (0) - 1} \Gamma[- \alpha_{v} (0)].
\end{align}

To numerically evaluate the differential cross section, the nucleon form factor $A(t)$ needs to be specified.
For this we employ the gravitational form factor calculated with the bottom-up AdS/QCD model~\cite{Abidin:2009hr}.
Utilizing the soft-wall model, in which the AdS geometry is smoothly cut off in the infrared region, it can be obtained as
\be
A (t) = (a + 1) \left[ - ( 1 + a + 2 a^2 ) + 2 a ( 1 + 2 a^2 ) \Phi ( -1, 1, a ) \right],
\ee
where $\Phi ( -1, 1, a )$ is the LerchPhi function, and $a = t / (8 \kappa^2)$ in which $\kappa$ is a parameter.
$\kappa$ affects the behavior of the nucleon wave function at the infrared boundary, and can be determined with the nucleon and $\rho$ meson mass, $m_n$ and $m_\rho$, following the relations: $m_n^2 = 8 \kappa^2$ and $m_\rho^2 = 4 \kappa^2$.
Hence, this form factor does not bring any adjustable parameters into the present model setup.

%%%%%%%%%%%%%%%%%%%%%%%%%%%%%%
\section{\label{sec:numerical_results}Numerical results}
%%%%%%%%%%%%%%%%%%%%%%%%%%%%%%
Here we explain how to determine the model parameters and display the numerical results.
The present model involves eight parameters in total.
However, for most of them we can utilize the values determined in the preceding studies, due to the universality of the Pomeron and Reggeon.
All the parameter values used in this study are summarized in Table~\ref{table_for_parameters}.
\begin{table}[tb]
\centering
\caption{Parameter values.}
\begin{tabular}{l l l}
\hline
Parameter&~~~~Value&~~~~~~~Source\\                
\hline
$\alpha_g(0)$ &~~~~1.084 &~~~~~~~fit to $pp(p \bar{p})$ scattering data at high energies~\cite{Xie:2019soz} \\
$\alpha'_g$ &~~~~0.368~$\rm{GeV}^{-2}$ &~~~~~~~fit to $pp(p \bar{p})$ scattering data at high energies~\cite{Xie:2019soz} \\
$\lambda_{g}$ &~~~~9.593~$\rm{GeV}^{-1}$ &~~~~~~~fit to $pp(p \bar{p})$ scattering data at high energies~\cite{Xie:2019soz} \\
$\alpha_v(0)$ &~~~~0.444 &~~~~~~~fit to $pp(p \bar{p})$ scattering data at medium energies~\cite{Liu:2022zsa} \\
$\alpha'_v$ &~~~~0.925$~\rm{GeV}^{-2}$ &~~~~~~~fit to $pp(p \bar{p})$ scattering data at medium energies~\cite{Liu:2022zsa} \\
$\lambda_{vpp}$ &~~~~7.742~ &~~~~~~~fit to $pp(p \bar{p})$ scattering data at medium energies~\cite{Liu:2022zsa} \\
$\lambda_{v \bar{p} \bar{p}}$ &~~~~16.127 &~~~~~~~fit to $pp(p \bar{p})$ scattering data at medium energies~\cite{Liu:2022zsa} \\
$\lambda_{vnn}$ &~~~~8.088 $\pm~0.029$ &~~~~~~~this work \\
\hline
\end{tabular}
\label{table_for_parameters}
\end{table}
For the Pomeron related parameters, we utilize the values obtained in Ref.~\cite{Xie:2019soz}, in which only the Pomeron exchange contribution was taken into account to investigate the $pp$ and $p \bar{p}$ cross sections, focusing on the high energy region.
For the Reggeon related parameters, except for the Reggeon-neutron coupling $\lambda_{vnn}$, we utilize the values determined in Ref.~\cite{Liu:2022zsa}, in which both the Pomeron and Reggeon contributions were considered to describe the $pp$ and $p \bar{p}$ total cross sections in the medium energy region.
Since the charge difference affects the Reggeon coupling, the Reggeon-proton and Reggeon-antiproton coupling constant, $\lambda_{vpp}$ and $\lambda_{v \bar{p} \bar{p}}$, are different from each other.

\begin{figure}[tb!]
\begin{center}
\includegraphics[width=0.62\textwidth]{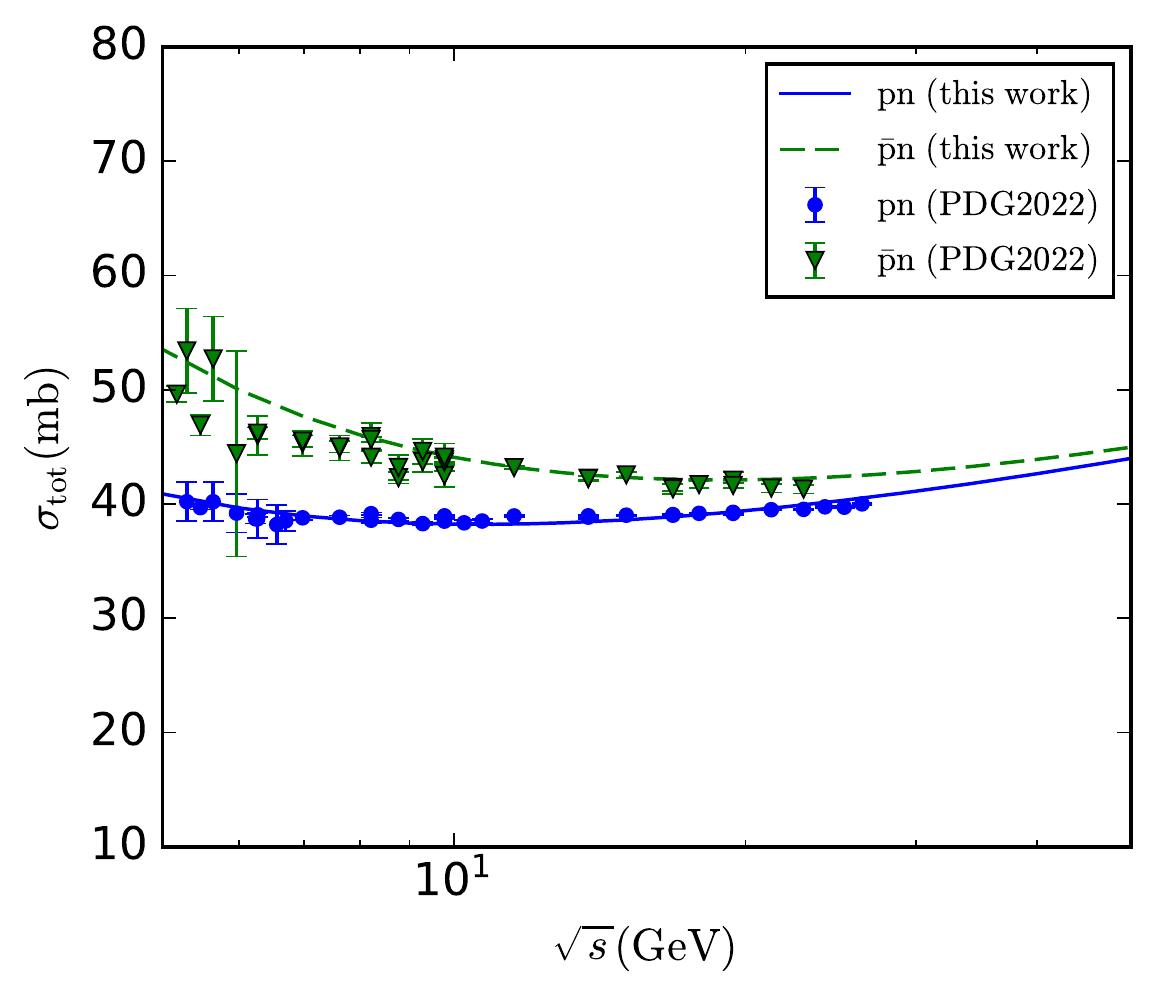}
\caption{
The total cross section as a function of $\sqrt{s}$.
The solid and dashed curve represent our calculations for the $pn$ and $\bar{p} n$ scattering, respectively.
The experimental data are taken from Ref.~\cite{ParticleDataGroup:2022pth} and depicted with error bars.
}
\label{fig1}
\end{center}
\end{figure}

\begin{figure}[tb!]
\begin{center}
\includegraphics[width=0.62\textwidth]{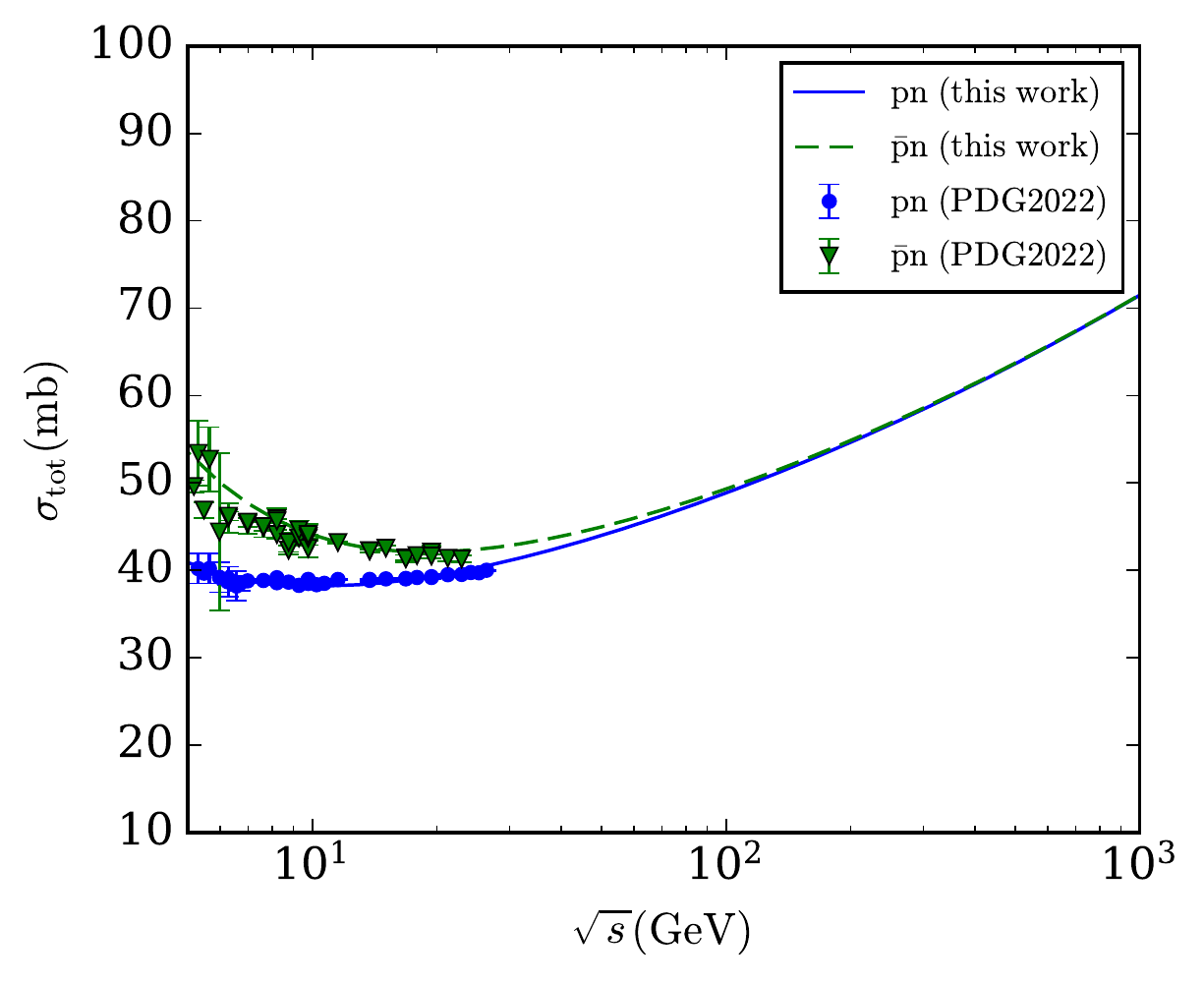}
\caption{
Similar to Fig.~\ref{fig1}, but for the wider $\sqrt{s}$ range.
}
\label{fig_added}
\end{center}
\end{figure}

Therefore, to investigate the $pn$ and $\bar{p} n$ scattering, we only need to determine a single adjustable parameter, which is the Reggeon-neutron coupling constant $\lambda_{vnn}$.
In this study, this determination is done by performing a numerical fit, utilizing the MINUIT package~\cite{James:1975dr} and focusing on the total cross section at $\sqrt{s} \geq 5$~GeV.
Using Eq.~\eqref{eq:tcs_final} and the $pn$ and $\bar{p} n$ total cross section data summarized by the Particle Data Group (PDG) in 2022~\cite{ParticleDataGroup:2022pth}, we determine $\lambda_{vnn}$, and the resulting best fit value is shown in Table~\ref{table_for_parameters}.
We display our calculations in Figs.~\ref{fig1} and \ref{fig_added}.
It is seen that the experimental data for both the $pn$ and $\bar{p} n$ scattering are well described with the present model in the medium energy region, where both the Pomeron and Reggeon exchange give substantial contributions to the cross section.

\begin{figure}[tb!]
\begin{center}
\includegraphics[width=0.7\textwidth]{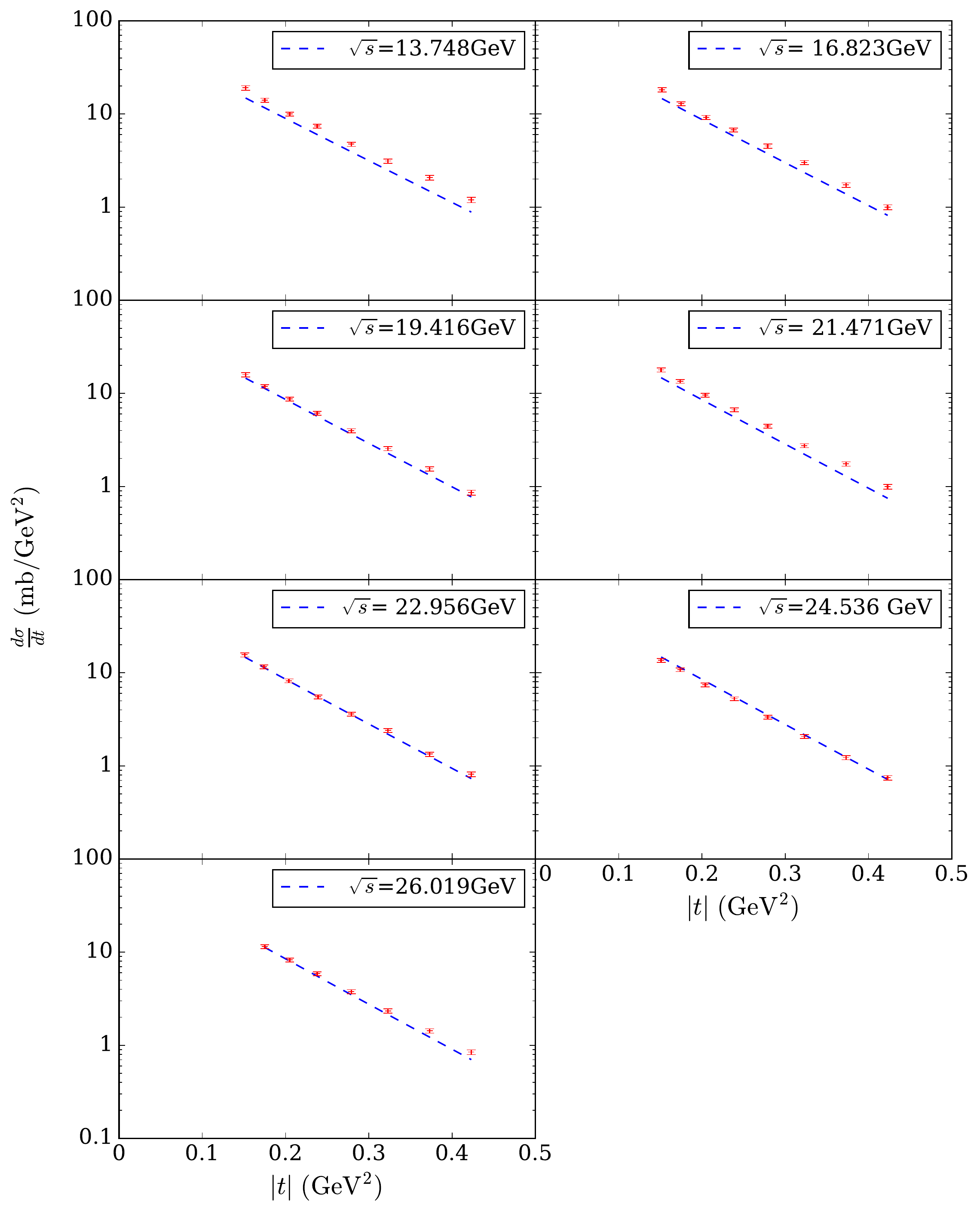}
\caption{
The differential cross section of the $pn$ scattering as a function of $|t|$ for various $\sqrt{s}$.
The dashed curves represent our predictions.
The experimental data are taken from Ref.~\cite{DeHaven:1978kj} and depicted with error bars.
}
\label{fig2}
\end{center}
\end{figure}

Once the parameter $\lambda_{vnn}$ is determined, using Eq.~\eqref{eq:dcs_final}, we can predict the differential cross section without any additional parameters.
Since we focus on the Regge regime and need to avoid the effect of the Coulomb interaction, similar to the previous works~\cite{Xie:2019soz,Liu:2022zsa}, we limit the considered kinematic region to $\sqrt{s} \geq 10$~GeV and $0.01 \leq |t| \leq 0.45$~GeV$^2$, and show our predictions for the $pn$ scattering in Fig.~\ref{fig2}.
Although the $\sqrt{s}$ range, in which the experimental data exist, is narrow, it is found that our predictions are overall consistent with the data in the considered kinematic region.
In the first two panels, there are some deviations between our predictions and the data, which may be due to the relatively low center-of-mass energies.
Some deviations are also seen in the result for $\sqrt{s} = 21.471$~GeV, but our results for $\sqrt{s} = 19.416$ and $22.956$~GeV are well.

\begin{figure}[tb!]
\begin{center}
\includegraphics[width=0.95\textwidth]{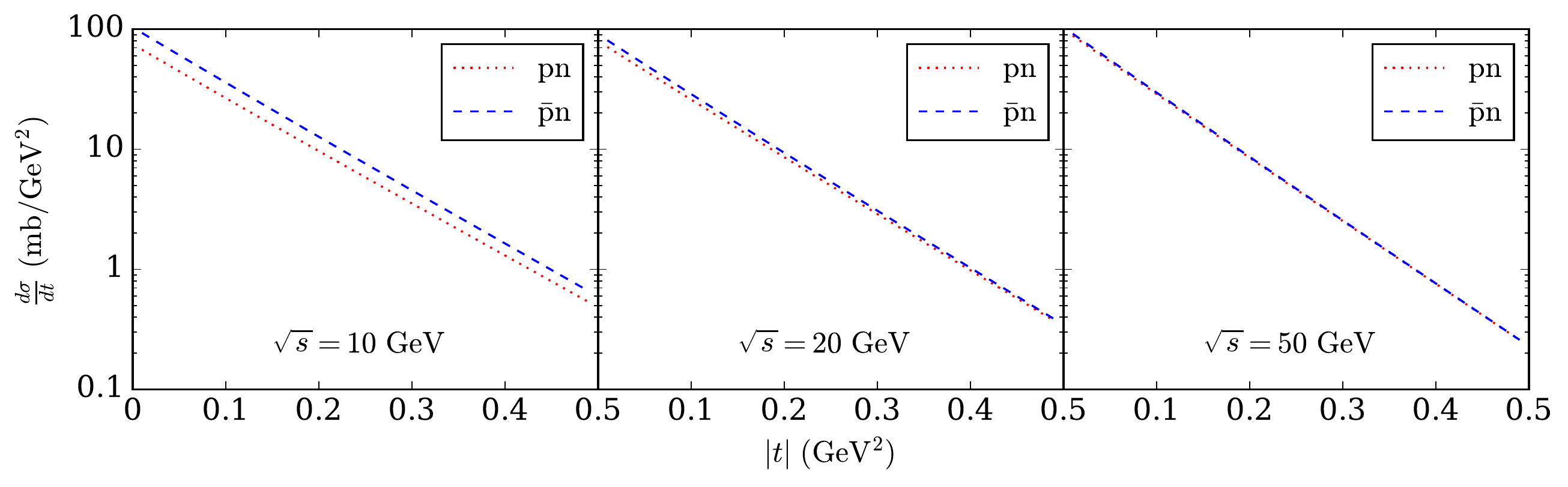}
\caption{
The differential cross section as a function of $|t|$ for $\sqrt{s} =$ 10, 20, and 50~GeV.
The dotted and dashed curves represent our predictions for the $pn$ and $\bar{p} n$ scattering, respectively.
}
\label{fig3}
\end{center}
\end{figure}

As to the $\bar{p} n$ scattering, currently there are no available data with which we can compare our predictions.
Hence we present the comparison between our predictions for the $pn$ and $\bar{p} n$ scattering in Fig.~\ref{fig3}.
It is seen that the differences between the $pn$ and $\bar{p} n$ results decrease as the energy increases, which reflects the fact that the Reggeon contribution decreases and the Pomeron contribution becomes dominant.
At $\sqrt{s} = 50$~GeV, the both results take the almost same values.
The energy dependence of the Pomeron and Reggeon exchange contribution was investigated in detail for the $pp(p \bar{p})$ scattering in Ref.~\cite{Liu:2022zsa}, and our predictions are consistent with their results.

%%%%%%%%%%%%%%%%%%%%%%%%%%%%%%
\section{\label{sec:conclusion}Conclusion}
%%%%%%%%%%%%%%%%%%%%%%%%%%%%%%
We have investigated the elastic $pn$ and $\bar{p} n$ scattering in a holographic QCD model, taking into account the Pomeron and Reggeon exchange in the Regge regime.
In our model setup, the Pomeron and Reggeon exchange are described by the Reggeized spin-2 glueball and vector meson propagator, respectively, and for the Pomeron-nucleon coupling the nucleon gravitational form factor, which can be calculated with the bottom-up AdS/QCD model, is utilized.
There are several parameters in the model, but for most of them we can employ the values determined in the preceding studies, due to the universality of the Pomeron and Reggeon.
Hence we only need to determine a single parameter, which is the Reggeon-neutron coupling constant, to calculate the $pn$ and $\bar{p} n$ cross sections.

Focusing on the total cross section, we have performed a numerical fit to determine this parameter, and found that the experimental data of both the $pn$ and $\bar{p} n$ can be well described within the model.
Once the parameter is determined, the differential cross section can be predicted without any adjustable parameters.
Although the currently available data are limited, we have explicitly shown that our predictions are consistent with those, which implies the predictive ability of the present model.
Further applications to other scattering processes are certainly needed.
Moreover, it is expected that future experiments of high energy forward scattering will help to better constrain the model and improve our understanding of the nonperturbative nature of the strong interaction, which is one of the most important problems in high energy physics.

%%%%%%%%%%%%%%%%%%%%%%%%%%%%%%
\section*{Acknowledgments}
%%%%%%%%%%%%%%%%%%%%%%%%%%%%%%
The work of A.W. was supported by the start-up funding from China Three Gorges University.
A.W. is also grateful for the support from the Chutian Scholar Program of Hubei Province.

\bibliography{references}

\end{document}